%% file: elsarticle-template-num.tex
\documentclass[final,5p,times,twocolumn,authoryear]{elsarticle}

\usepackage{amssymb}
\usepackage{amsmath}
\usepackage[british]{babel}
\usepackage{xcolor}
\usepackage[colorlinks=true]{hyperref}
\usepackage[inline]{enumitem}
\usepackage[dvipsnames]{xcolor}
\usepackage[ruled, linesnumbered]{algorithm2e}
\usepackage{csquotes}
\usepackage{makecell}
\usepackage{multirow}
\usepackage{caption}
\usepackage{subcaption}

\begin{document}

\begin{frontmatter}



\title{
AlertBERT: A noise-robust alert grouping framework for simultaneous cyber attacks
}


\author{Lukas Karner\corref{cor}} %% Author name
\ead{lukas.karner@ait.ac.at}
\cortext[cor]{Corresponding Author.}

\author{Max Landauer}
\ead{max.landauer@ait.ac.at}

\author{Markus Wurzenberger}
\ead{markus.wurzenberger@ait.ac.at}

\author{Florian Skopik}
\ead{florian.skopik@ait.ac.at}

\affiliation{organization={AIT Austrian Institute of Technology},%Department and Organization
            addressline={Giefinggasse 4}, 
            city={Vienna},
            postcode={1210}, 
            state={Vienna},
            country={Austria}}

%% Abstract
\begin{abstract}
Automated detection of cyber attacks is a critical capability to counteract the growing volume and sophistication of cyber attacks.
However, the high numbers of security alerts issued by intrusion detection systems lead to alert fatigue among analysts working in security operations centres (SOC), which in turn causes slow reaction time and incorrect decision making. 
\textit{Alert grouping}, which refers to clustering of security alerts according to their underlying causes, can significantly reduce the number of distinct items analysts have to consider.
Unfortunately, conventional time-based alert grouping solutions are unsuitable for large scale computer networks characterised by high levels of false positive alerts and simultaneously occurring attacks.
To address these limitations, we propose \textit{AlertBERT}, a self-supervised framework designed to group alerts from isolated or concurrent attacks in noisy environments.
Thereby, our open-source implementation of AlertBERT leverages masked-language-models and density-based clustering to support both real-time or forensic operation.
To evaluate our framework, we further introduce a novel data augmentation method that enables flexible control over noise levels and simulates concurrent attack occurrences.
Based on the data sets generated through this method, we demonstrate that AlertBERT consistently outperforms conventional time-based grouping techniques, achieving superior accuracy in identifying correct alert groups.
\end{abstract}


%Research highlights
% \begin{highlights}
% \item AlertBERT groups alerts robustly in high-noise data with simultaneous cyber attacks.
% \item AlertBERT is self-supervised, using masked-language-models and clustering methods.
% \item Data augmentation increases noise levels and cyber attacks in alert datasets.
% \item AlertBERT outperforms conventional time-based methods in large computer networks.
% \end{highlights}

%% Keywords
\begin{keyword}
cyber security \sep log data analysis \sep alert grouping \sep machine learning \sep natural language processing \sep clustering


\end{keyword}

\end{frontmatter}



\include{contents}

\appendix

%%       \cite{<label>} ==> [1]


\bibliographystyle{elsarticle-harv} 
\bibliography{literature}


\end{document}

%% file: contents.tex
\section{Introduction}
\label{sec:intro}

In modern computer networks, \textit{intrusion detection systems} (IDS) play an essential role in cyber security operations \citep{99falsepos}.
The alerts generated by these systems are a key component of continuous efforts to defend networks against hostile attacks and ensure their continued operation.
However, cyber security analysts are increasingly confronted with a growing volume of alerts that must be assessed for validity and urgency.
This phenomenon, commonly referred to as \textit{alert flooding} \citep{Landauer2022Dealing, eckhoff_neurosymbolic}, can ultimately lead to \textit{alert fatigue} \citep{fatigue2, eckhoff_neurosymbolic}, meaning the desensitisation of analysts towards potential threats and wrong decisions in handling alerts caused by this.

\textit{Alert grouping} is the task of aggregating alerts according to their underlying causes.
As single attack events usually cause an IDS to issue multiple alerts, for example, network scans can cause an IDS to issue many thousands of alerts \citep{aitads}, alert grouping poses a high potential to alleviate the workload of cyber security analysts.
Hence, an alert grouping system can significantly improve the security operations of such a network, because it allows security analysts to identify attacks more quickly and accurately and to focus on taking countermeasures against the attacks.
Furthermore, the insights provided by an alert grouping system can be used for other downstream cyber security operations tasks.
For example, alert grouping can be used to remove redundant alerts, it is a necessary prerequisite for the generation of meta-alerts used for recognition of previous attack occurrences \citep{Landauer2022Dealing}, it can be used for the detection of false positives in the alerts \citep{metrics}, or the identified alert groups can be employed for attack chain modelling \citep{eckhoff_graph}.

State-of-the-art publicly available solutions to the alert grouping problem, e.g.\ the time-delta method proposed by \citep{Landauer2022Dealing}, already produce robust results in deployment scenarios of medium-sized enterprise networks with little background noise\footnote{We define as \textit{noise} any alerts that are not caused by an actual attack on the computer system, i.e.\ false positive alerts.} in the alerts.
In scenarios with large computer networks, entailing high levels of background noise and potentially multiple attacks occurring at the same time, however, the problem of alert grouping still has no satisfying solution.
This is due to the fact, that the alert sequences produced by IDSs in such cases have several properties which cause the quality of existing alert grouping methods to degrade \citep{Landauer2022Dealing}.
These properties are, for example, a strongly varying density of alerts over time, a too high density of alerts in time, or alerts that occur without sufficient time gap to enable separation by purely time-based methods.

Strongly varying alert densities can be problematic for alert grouping methods, because these methods might rely on the alert density as an indicator of potentially ongoing attacks \citep{bayesianblocks}.
If the alert density varies over time in ways which are not differentiable from the patterns associated with attacks or are otherwise irregular, then the alert grouping methods might have problems in adapting their operations under these circumstances.
Permanently high alert densities, either caused by high levels of background noise in the alerts or simply by a large number of devices in the computer network, on the other hand, can be problematic for alert grouping systems as well.
This is because the variations in the alert density which are used by alert grouping systems can be drowned out in high noise levels, for example, when unusual spikes of alerts are concealed by the overall noise level fluctuations, again disrupting their operations.
Finally, alert groups which are not separable in time are problematic for alert grouping systems which rely on the temporal continuity of alert groups for their operation.
Such alert grouping systems can, for example, be confounded by two simultaneous attacks occurring in the network which produce alert sequences that overlap in time and, therefore, cannot be assigned to different attacks by the system.

Thus, we introduce the novel AlertBERT framework to tackle the alert grouping problem in such challenging environments.
Our contributions are:
A theoretical framework to solve the alert grouping problem in a self-supervised manner based on masked-language-models and density-based clustering;
as well as a data augmentation method to enhance alert datasets by increasing the level of noise present in them and modify the occurrence of attacks.
The code implementing our prototype of the AlertBERT framework and its evaluation is publicly available.\footnote{\url{https://github.com/ait-aecid/AlertBERT}}

The remainder of this paper is structured as follows:
Section \ref{sec:lit} discusses further literature related to our work; 
Section \ref{sec:method} introduces our proposed AlertBERT framework for alert grouping; 
in Section \ref{sec:eval} we present the methods used to evaluate our framework and the results of this evaluation;
Section \ref{sec:dis} discusses these results and their implications;
and Section \ref{sec:con} concludes this paper.

\begin{figure*}[t]
    \includegraphics[width=\linewidth]{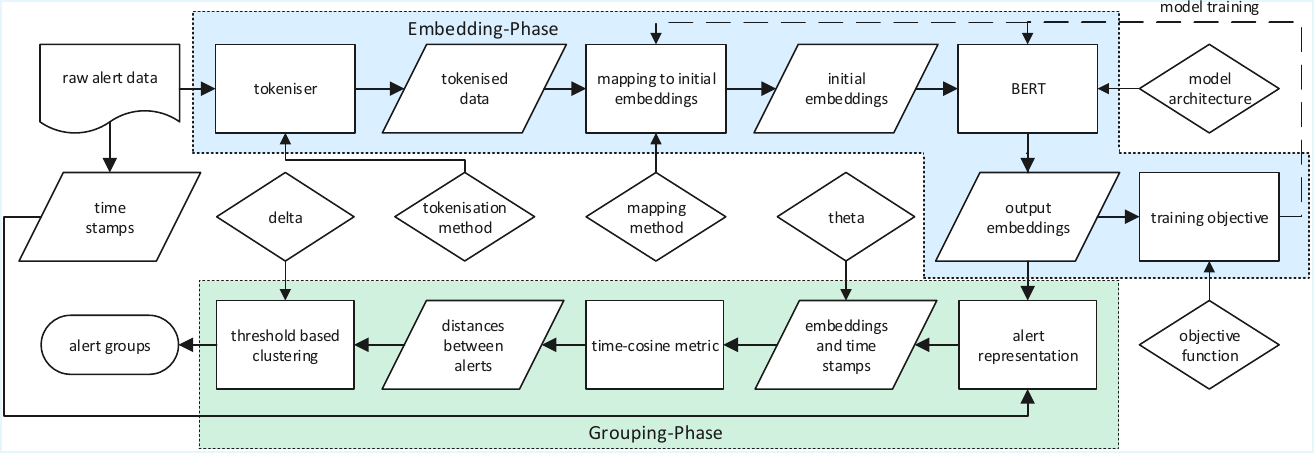}
    \caption{
        An illustration of the AlertBERT framework showing the components of the framework, the data flow among them, and the parameters to set in them.
        For each of the two phases of the framework, the components belonging to them and their internal data states are highlighted.
    }
    \label{fig:flowchart}
\end{figure*}

\section{Related Work}
\label{sec:lit}

To the best of our knowledge, the following methods for alert grouping have been proposed in the literature, however each of the proposed methods is based on different assumptions and requirements.
Thus, there is a large variety in the properties that these methods have and, subsequently, also in their applicability to various application scenarios.

The time-delta method \citep{Landauer2022Dealing} is based purely on timestamps of alerts and considers two alerts to be part of the same alert group if the difference between their timestamps is below a given threshold.
Due to its simplicity and its outstanding effectiveness in predictable and noise-free environments, the time-delta method is very versatile and light-weight tool for security operations in small to medium sized networks.
However, because of its dependence on timestamps, time-delta is sensitive to high noise levels and it is not able to resolve alert groups that overlap in time.
Thus, despite its advantages, time-delta is not suitable for large-scale security operations.

The approach proposed by \citep{eckhoff_graph} uses timestamps and attributes of alerts to create a graph structure capturing the relations among them.
Through temporal thresholds this global graph is partitioned to define alert groups as local graphs and, furthermore, these local graphs are used as fingerprints of attacks to correlate attacks by graph-matching methods.
While the idea behind the alert grouping approach of \citep{eckhoff_graph} appears promising because it is an unsupervised method, the fact that the method uses one hyper-parameter for each alert attribute it takes into account can make it prohibitively costly to configure in practice.
For example, alerts produced by one IDS can have more than 20 distinct attributes, and it is common that multiple IDSs are deployed in one computer system, driving up the number of distinct attributes to consider \citep{aitads}.

The framework proposed by \citep{metrics} create alert groupings through incremental clustering applied to alert embeddings which are derived from different pre-defined alert similarity metrics and an embedding model trained on them.
A disadvantage of this approach is the manual definition of similarity measures for different types of alerts that is required there.
Thus, its configuration and continued operation pose a significant effort hindering the utilisation of this method in large computer networks due to the high number of alert types encountered there.

The method presented by \citep{clusters14} employs a pre-processing pipeline for alert messages with hard-coded feature extraction and similarity metrics to define alert groups through hierarchical clustering.
The pre-processing and feature extraction of alert messages used in this method are specific to the alerts generated by a specific computer system and, furthermore, the method can only be applied for posterior analyses of recorded alerts because its clustering results cannot be updated in an incremental way.
Thus, also this method is not suitable to be employed in large scale cyber security operations because of the lack of flexibility in its configuration and, additionally, it may not be used for real-time alert monitoring anyway.

Apart from methods that focus purely on alert grouping there is also the adjacent field of alert aggregation techniques that aim to extract information from alerts and present it in a more compact form.
Among these there are, for example, methods to create meta-alerts from alert graphs \citep{aggregate1, aggregate2}, with the help of pattern mining \citep{aggregate3}, or through shared attributes \citep{aggregate4}.
Other than that, there are alert aggregation methods that focus on attack classification and do so by considering various attributes of alerts or combinations thereof, such as IP addresses \citep{aggregate5, aggregate6}, parts of IP addresses and ports \citep{aggregate7}, or combinations of attributes and timestamps \citep{aggregate8}. 

Altogether, we find the existing alert grouping methods to be unsuited for applications with high noise levels and simultaneous attacks, which are common in large scale computer networks.
Therefore, it is necessary to develop an alert grouping framework able to fill this gap, and we do so through our proposed method.

\section{Alert Grouping Framework}
\label{sec:method}

This section introduces our alert grouping framework by, first, providing a high-level overview of its two main parts and, then, explaining the two parts of it in detail.

\subsection{Overview}

To solve the alert grouping problem in environments with high-noise levels and potentially simultaneous attacks, we will employ a combination of \textit{machine learning} techniques, specifically from the domains of \textit{natural language processing} and \textit{density-based clustering}.
To this end, the AlertBERT framework consists of two main components which we will discuss in detail:
\begin{enumerate}
    \item \label{it:emb} Embedding-Phase:
    The goal of this phase is to create numerical representations of alerts which are rich enough to capture the information relevant for alert grouping.
    This is done by means of masked-language-models and self-supervised learning methods equipped with our adaptions for making them suitable to process alerts.
    In this way, our framework leverages the flexibility of learned representations that need minimal human supervision to find alert embeddings suitable for the grouping problem.
    \item \label{it:grp} Grouping-Phase:
    This phase is concerned with extracting alert groups from the alert embeddings obtained in the first phase.
    This is done by considering the distances in time and embedding dimensions of the individual alert representations and applying density based clustering to them.
    Like this, the AlertBERT framework naturally extends the simplicity and robustness of the time-delta method \citep{Landauer2022Dealing} to the high-noise and simultaneous-attack case, thus enabling robust and accurate alert grouping also in these cases.
\end{enumerate}

The AlertBERT framework is, thus, suitable to be employed in real-world applications for multiple reasons:
\begin{enumerate*}[label=(\roman*)]
    \item The additional embedding dimensions employed by the framework make the grouping noise-robust as they allow to avoid collisions among the alert representations in case of high alert density in time.
    \item Similarly, the embedding dimensions facilitate the separation of simultaneous attacks in the alert grouping as they allow for different alert groups that overlap in time.
    \item The frameworks self-supervised training does not require access to true alert groupings that are unfeasible to obtain in practice.
    \item The framework allows for flexible operation as it can be used either in an online\footnote{Incremental processing of an incoming stream of samples.} or forensic manner.
    \item The framework can handle alerts of unknown formats without need for human intervention (defining schemata, etc.) as masked language models with our adaptions can dynamically adjust to the contents of alerts.
\end{enumerate*}

An overview of the phases of the AlertBERT framework, their components, the data flow among them, and their parameters is provided in Figure \ref{fig:flowchart}.

\subsection{Embedding-Phase}
\label{sec:emb}

The embedding-phase of the AlertBERT framework is concerned with the numerical representation of IDS alerts in the form of embeddings and pursues the objective of preserving as much information as possible about the alerts, their structure, and the relationships among them.
In particular, to obtain embeddings useful for alert grouping, this means that alerts that are caused by the same hostile activity against the network should be represented by similar embeddings so that the grouping-phase of the framework can exploit this similarity to reconstruct their common origin.

With this requirement on the nature of the embeddings and the additional practical constraint of self-supervision, \textit{BERT} models \citep{devlin-etal-2019-bert}, i.e.\ transformer-encoder neural networks trained on a masked-language objective, are a natural choice for this application.
This is because, on the one hand, BERT models are the most well-established model architecture for text-based data due to their ease of use and low training complexity in contrast to more resource intensive \textit{large-language-models} \citep{modernbert}.
On the other hand, the embeddings produced by BERT models are well-known for their semantic usefulness \citep{beyondwords, clark-etal-2019-bert} and applicability to downstream tasks such as clustering or information retrieval \citep{Gardazi2025, bertir}.
As IDS alerts generally are a semi-structured text-based form of data, e.g.\ a temporal sequence of JSON objects, they are suitable to be processed with BERT models and, thus, we employ them to form the foundation of the embedding-phase of the AlertBERT framework.

The abstract template for our alert embedding model is, thus, given as a BERT model with our contributions being the modifications to adapt it to the alert grouping use case via appropriate \begin{enumerate*}[label=(\roman*)]
    \item \label{it:tok} alert tokenisation, 
    \item \label{it:map} mapping to initial embeddings, and 
    \item \label{it:obj} a training objective suitable for these.
\end{enumerate*}
We will now discuss these steps individually in the following subsections.

\subsubsection{Alert Tokenisation}

\begin{figure}[t]
    \frenchspacing
    \texttt{\{ "time": "2025-08-10T13:42:07.391",\\
    \hspace*{1em}"ip": "10.35.35.204",\\
    \hspace*{1em}"message": "AMiner: CPU value deviates\\
    \hspace*{2em}from average in monitoring logs.",\\
    \hspace*{1em}"type": "Statistical data report", \dots\}}
    \caption{A simplified example of an alert in JSON-format produced by the AMiner \citep{aminer} IDS as it is part of the AIT Alert Dataset \citep{aitads}.}
    \label{fig:alert}
\end{figure}

For alert tokenisation we heavily rely on the semi-structured nature of the alerts to extract features from them which are useful for alert grouping, an example of such an alert can be seen in Figure \ref{fig:alert}.
In the most basic case, an alert can simply be tokenised as a set of hard-coded features of this alert where the features are indicative of the relations among the alerts and share values across them.
Examples of such features could be the type of alert, the device from which the alert originated, or the IP address the connection to which caused the alert.
In our prototype implementation of the AlertBERT framework, we employed the alert type and the originating device as hard-coded features for alert tokenisation as these features can be easily extracted from the alerts of the AIT Alert dataset, they do not contain missing values, and are sufficiently expressive for prototyping purposes.

More advanced tokenisation methods may not rely on hard-coded features of the alerts any more, but may instead exploit the internal structure of the alerts in purely textual form or based on the tree-like graph structure inherent in them, e.g.\ the JSON format.
When interpreting alerts in their purely textual form, the usual tokenisation techniques of natural language processing, e.g.\ byte pair encoding \citep{radford2019language, zhuang-etal-2021-robustly}, can be applied to them, while the interpretation of alerts based on graph structures requires more nuanced tokenisation methods, which take into account both the graph structure and the textual or numeric nature of the alert's attributes, such as \citep{jsongrindr}.

\subsubsection{Mapping to Initial Embeddings}

To fully understand the aspects of the alert tokenisation methods, one has to consider also the respective mapping to initial embeddings used along with each tokenisation method as they are closely linked to each other.
Generally, we define a mapping to initial embeddings on the alert level, i.e.\ as a function taking as input the token structure in which an alert was decomposed in step \ref{it:tok}, assigning a learnable numeric representation to each of the tokens, and then aggregating all of the token representations into a single embedding vector representing the entire alert.

For the basic tokenisation technique of just using hard-coded features, the mapping to initial embeddings can be implemented as a sum of feature embeddings where each feature is associated with a trainable vector and the initial embedding of an alert is defined as the sum of these vectors corresponding to the features of this alert.
In our prototype implementation of the AlertBERT framework, with alert type and originating device as hard-coded features for alert tokenisation, we apply the sum-of-features method as mapping to initial embeddings.
More precisely this means that any possible value\footnote{
Rare values, outliers, or values not present in the training data may be represented by an \textquote{unknown} token.
} of the two features gets assigned a trainable vector as feature embedding, and the initial embedding of an alert is given by the sum of the feature embedding vectors belonging to the feature values of the alert.

In the more advanced cases of tokenisation techniques the mapping to initial embeddings can, for example, have the following implementations:
If the alerts are tokenised as texts, the token sequence corresponding to an alert can be passed on to a recurrent model such as an LSTM \citep{lstm} which accumulates it into a hidden state that serves as the initial embedding of this alert for the BERT model.
Similarly, in the case that the alerts are tokenised based on their graph structure, the vectors that were assigned to the nodes and edges of the graph can be iteratively accumulated by a graph-neural-network \citep{gnn} into a single embedding representing the entire graph.

Due to its particular importance for the alert grouping problem, we point out that also the timestamp of an alert can be incorporated into the alert's embedding as it provides valuable information about the order and temporal density of the alerts and can thus be seen as a generalisation of the positional encoding used in purely textual BERT models.
This can happen either right at the initial embedding stage, or, depending on the specific BERT model used, it can also happen at a later stage inside the model, e.g.\ via rotary positional encoding \citep{rope}.

The entire process of alert tokenisation and mapping to initial embeddings is summarised in Algorithm \ref{alg:tokmap}.
There, looping over all alerts given in the input data, line \ref{lin:a} extracts the token structure of an alert, line \ref{lin:b} uses the structure to tokenise the alert, line \ref{lin:c} assigns the token-level embeddings to each token, \ref{lin:d} aggregates the token-embeddings into the alert's initial embedding according to the token structure, and line \ref{lin:e} appends the alert's initial embedding to the output sequence of initial alert embeddings.
Together, lines \ref{lin:a} and \ref{lin:b} correspond to the alert tokenisation~\ref{it:tok}, and lines \ref{lin:c} and \ref{lin:d} correspond to the mapping to initial embeddings \ref{it:map}.
\begin{algorithm}[t]
    \caption{Alert Tokenisation and Mapping to Initial Embeddings}
    \label{alg:tokmap}
    \SetKwFunction{TokenStructure}{TokenStructure}
    \SetKwFunction{Tokeniser}{Tokeniser}
    \SetKwFunction{TokenEmbedding}{TokenEmbedding}
    \SetKwFunction{InitialEmbedding}{InitialEmbedding}
    \SetKwFunction{Length}{Length}

    \KwData{Sequence $A$ of alert objects.}
    \KwResult{Sequence $E$ of alert embeddings.}
    \BlankLine
    $E=()$\;
    \For{$i=1$ \KwTo \Length{$A$}}{
        $S=$ \TokenStructure{$A[i]$}\label{lin:a}\;
        $T=$ \Tokeniser{$A[i]$, $S$}\label{lin:b}\;
        $T=$ \TokenEmbedding{$T$}\label{lin:c}\;
        $e=$ \InitialEmbedding{$T$, $S$}\label{lin:d}\;
        $E=E+(e)$\label{lin:e}\;
    }
\end{algorithm}

\subsubsection{Training Objective}

Finally, the training objective used to define how the BERT model shall reconstruct the sequence of initial embeddings provided to it is what ultimately determines which aspects of the alerts will be represented in the embeddings.
The choice of this objective ultimately depends on the tokenisation and embedding methods used in the embedding and grouping-phases, but, still, it should also reflect which properties and attributes of alerts are considered useful for the specific alert grouping problem at hand.

In the case of alert tokenisation by hard-coded features, the training objective can simply be to predict these features of the masked alerts.
Likewise, in our prototype implementation of the AlertBERT framework where we use the alert type and originating device as hard-coded features for alert tokenisation, we let a classification head predict the values of these two features based on the output embeddings of the masked-language-model for its training.

In the case of the more advanced alert tokenisation methods, which incorporate the entire structure and content of the alert, also more advanced self-supervised training objectives can be used to obtain informative alert embeddings.
Such objective can be, for example, self-distillation \citep{dino}, joint-embedding-prediction \citep{Assran_2023_CVPR, tijepa}, or self-supervised sentence representation methods \citep{wu-zhao-2022-sentence, Wang_He_Wang_Zhou_Sun_Qiu_2024}.

\subsection{Grouping-Phase}
\label{sec:grp}

The grouping-phase of the AlertBERT framework, concerned with creating alert groups from the alert embeddings obtained in the embedding-phase of the framework, is inspired by the time-delta alert grouping method \citep{Landauer2022Dealing}.
Essentially, the time-delta method is a simple density-based clustering \citep{cluster} of the alerts, which assigns two alerts to the same group if they occurred within a given threshold of $\delta$ seconds, and it yields excellent state-of-the-art results in scenarios with medium-sized computer networks and little noise in the data \citep{Landauer2022Dealing}.
But, as outlined in Section \ref{sec:intro}, the time-delta method fails in scenarios with high numbers of noise alerts polluting the data or multiple attacks occurring at once.
This is a known limitation of the time-delta method already discussed in \citep{Landauer2022Dealing} as, once the alerts of different origins occur with a sufficient volume, the time dimension becomes saturated with them and the simple grouping rule of the time-delta method cannot separate them any more.
In other words, alerts of different origins tend to form large groups since noise alerts act as bridges between alerts that would otherwise be correctly separated into smaller groups.

Despite these issues, as density-based clustering is a versatile and self-supervised method, with some adaptions it is suitable to be employed as grouping method for the AlertBERT framework.
To mitigate the problems of the time-delta method, the grouping rule of AlertBERT does not only consider the timestamps of alerts, but it incorporates the alert embeddings obtained in the embedding-phase of the framework as additional dimensions in it.
In this way, alert groups can be separated in the embedding dimensions where it is not possible to do so in the time dimension alone and, therefore, the AlertBERT grouping method mitigates the issues that previous methods encounter in high-noise and simultaneous-attack scenarios.

Our alert grouping model is, thus, defined as a distance threshold based clustering, where two alerts are assigned to the same group if their distance is less than a given threshold value $\delta$, with our contributions being
\begin{enumerate*}[label=(\roman*)]
    \setcounter{enumi}{3}
    \item \label{it:rep} embedding-enriched alert representations, and
    \item \label{it:tcm} the so-called \textit{time-cosine metric} by which we determine the distances between the alerts.
\end{enumerate*}
We will now discuss these steps individually in the following subsections.

\subsubsection{Alert Representation}
\label{sec:rep}

In the grouping method of the AlertBERT framework, alerts $a_i$ are not only represented by their timestamp $t_i$.
Instead, as shown in Equation \ref{eq:alert}, the embeddings $e_i$ obtained from the embedding-phase of AlertBERT are appended to the timestamps to create alert representations, which not only respect the temporal distribution of the alerts, but also the information which was accumulated in the embeddings.
\begin{equation}
\label{eq:alert}
    a_i=(t_i,e_i)
\end{equation}
The alert embeddings used here will generally be the output embeddings of a BERT model trained according to the embedding-phase of our framework as these contain the largest amount of contextual information about other alerts occurring along with them, but it is also possible to extract embeddings at the initial or intermediate layers of the model.
While the two parts of the AlertBERT framework have been developed to be used in conjunction with each other, in principle, the grouping-phase works independently of the used alert embedding method and hence can also be used in combination with other frameworks, e.g.\ the alert vectors proposed by \citep{eckhoff_graph}.

\subsubsection{Time-Cosine Metric}

To assign the alerts to groups based on how close their representations are to each other, the AlertBERT framework defines a distance function for alert representations which takes into account both their temporal distance as well as the similarity of their embeddings.
This distance function we call the \textit{time-cosine metric} and for two alert representations $a_i=(t_i,e_i)$ for $i=1,2$ with timestamps $t_i$ and embeddings $e_i$ it is defined, as in Equation \ref{eq:metric}, as the maximum of their time difference and the cosine distance of their embeddings multiplied by a tuning factor $\theta$ which allows to balance the influences of the two components.
\begin{equation}
\label{eq:metric}
    d(a_1,a_2,\theta)=\max\left[|t_1-t_2|,\theta\cdot\left(1-\frac{\langle e_1|e_2 \rangle}{\Vert e_1\Vert\,\Vert e_2\Vert}\right)\right]
\end{equation}
As the time-cosine metric is defined as the maximum of two other metrics it is also a metric again itself.

While the time difference of alerts is the distance function used by the time-delta method \citep{Landauer2022Dealing}, the cosine distance is a common distance function to be used with embeddings produced by BERT models.
Thus, the combination of these two allows us to retain the excellent alert grouping efficacy of the time-delta method for simple scenarios while generalising it in such a way that it can also separate alert groups in scenarios with high levels of noise in the data and simultaneously occurring attacks.

\begin{figure}[t]
    \centering
    \includegraphics[width=\linewidth]{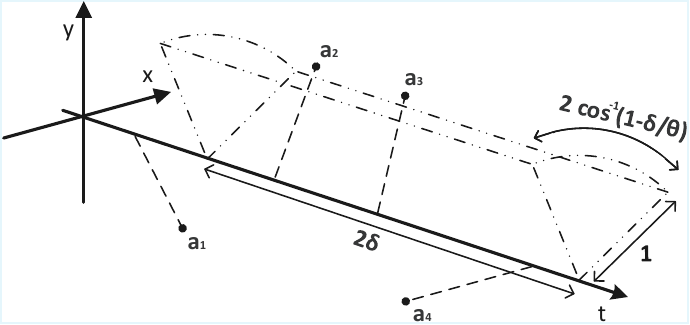}
    \caption{
        An illustration of a neighbourhood around an alert defined by the time-cosine-metric in the alert representation space.
    }
    \label{fig:metric}
\end{figure}

Figure \ref{fig:metric} provides an example of the neighbourhoods in the alert representation space defined by the time-cosine metric.
The graphic shows 4 alert representations $a_1,\dots,a_4$, each having 2 embedding dimensions denoted by the axes $x$ and $y$, their time dimension is shown along the axis $t$, and the dashed lines indicate the projections of the alert representations on the time axis.
Furthermore, the dash-dotted lines indicate the neighbourhood around the third alert representation $a_3$ defined by the time-cosine-metric with distance threshold $\delta$ and tuning factor $\theta$.
In the subspace defined by the embedding dimensions, the shape of this neighbourhood is solely defined by the cosine-similarity component of the time-cosine-metric and takes the shape of an unbounded cone centred at $a_3$.
The angular opening of the cone is equivalent to $2$ times a cosine-similarity of $\delta/\theta$, meaning that the opening angle has a size of $2\cos^{-1}(1-\delta/\theta)$~rad.
In time, the neighbourhood simply takes the shape of an interval of length $2\delta$, also centred at $a_3$.
In this example only alert $a_2$ would be assigned to the same group as alert $a_3$ because it lies inside the example neighbourhood while alert $a_1$ lies outside the neighbourhood in all dimensions and alert $a_4$ lies outside the neighbourhood in the embedding dimensions (but not in the time dimension).

\subsubsection{Grouping Algorithm}

Algorithm \ref{alg:grouping} provides an example implementation of the grouping-phase of the AlertBERT framework.
This algorithm determines the alert grouping by first constructing an edge list of the graph defined by the alert groups, and then passing the edge list to a function finding the connected components of this graph, which are just the alert groups.
More precisely, in the nested loops spanning lines \ref{lin:g} to \ref{lin:m} we iterate over all pairs of alert representations, in line \ref{lin:h} constructing the alert representation of the current alert, looping over all previous alerts in lines \ref{lin:i} to \ref{lin:m}, constructing their alert representation in line \ref{lin:j}, and checking whether the two alert representations have distance less than the threshold value $\delta$ according to the time-cosine-metric in line \ref{lin:k}.
If this is the case, then the pair of alert indices is appended to the edge list in line \ref{lin:l}.
Finally, in line \ref{lin:n} a function to compute the connected components is called on the edge list, which yields the alert group labels as result.
\begin{algorithm}[t]
    \caption{AlertBERT Grouping}
    \label{alg:grouping}

    \SetKwFunction{TimeCosineMetric}{TimeCosineMetric}
    \SetKwFunction{ConnectedComponents}{ConnectedComponents}
    \SetKwFunction{edges}{edge\_list}
    \SetKwFunction{nnodes}{number\_of\_nodes}

    \KwIn{Distance threshold $\delta$ and tuning factor $\theta$.}
    \KwData{Sequence $E$ of alert embeddings and sequence $T$ of timestamps.}
    \KwResult{Sequence $G$ of group labels.}
    \BlankLine
    $C=()$\label{lin:f}\; % \tcc*{initilise edge list}
    % $n=$ \Length{$E$}\;
    \For{$i=1$ \KwTo \Length{$E$}\label{lin:g}}{
        $a_1=(T[i],E[i])$\label{lin:h}\;
        \For{$j=1$ \KwTo $i-1$\label{lin:i}}{
            $a_2=(T[j],E[j])$\label{lin:j}\;
            \uIf{\TimeCosineMetric{$a_1$,$a_2$,$\theta$}$<\delta$\label{lin:k}}{
                $C=C+((i,j))$\label{lin:l}
            }
        }
    }\label{lin:m}
    $G=$ \ConnectedComponents{\label{lin:n}\\\quad\nnodes $=$ \Length{$E$},\\\quad\edges$=C$}\;
\end{algorithm}

Importantly, Algorithm \ref{alg:grouping} can be optimised by letting the inner loop (lines \ref{lin:i} to \ref{lin:m}) not iterate over all previous alerts, but only over those alerts which occurred at most $\delta$ seconds before the current alert as only these alerts can have a time-cosine-distance below the threshold $\delta$ to the current alert.

\section{Evaluation}
\label{sec:eval}

To evaluate the AlertBERT framework we compare its grouping outcomes to those of the state-of-the-art time-delta \citep{Landauer2022Dealing} method.
The details of this process are described in the following sections.

\begin{figure*}[t]
    \centering
    \includegraphics[width=\linewidth]{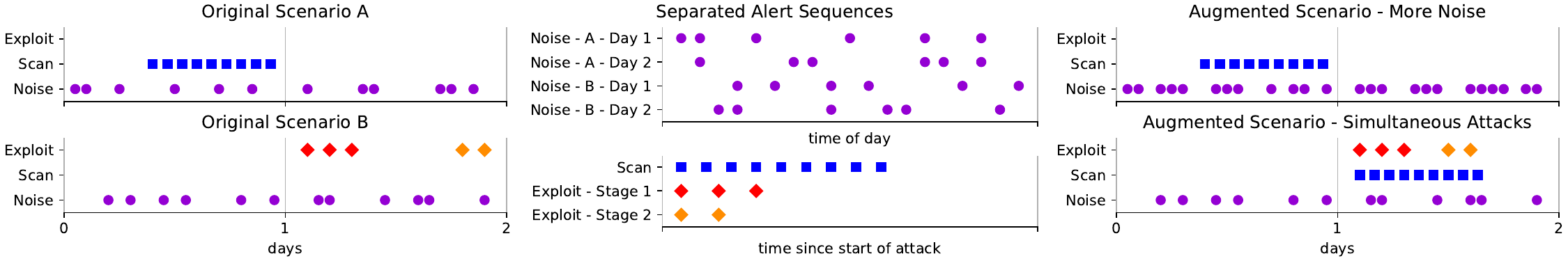}
    \caption{An illustration of our data augmentation method for creating AIT-ADS-A.}
    \label{fig:augmentation}
\end{figure*}

\subsection{Data}

We based the evaluation of the AlertBERT framework on the \textit{AIT Alert Dataset} (AIT-ADS) \citep{aitads} as it is, to the best of our knowledge, the only publicly available alert dataset that offers alert group labels.
This dataset was created by, first, launching a multi-step attack on a testbed representing a small enterprise network and, second, collecting the logs resulting from this \citep{logs}.
Third, the log data obtained in this way were then analysed by IDSs, and, fourth, the alerts produced by these were labelled according to their true origins\footnote{
In AIT-ADS these labels are called \textit{event labels}.
}.

In total, AIT-ADS consists of approximately 2.4 million alerts belonging to 8 recordings of alert streams obtained from different simulated computer networks, which we call \textit{scenarios}.
Each of the 8 scenarios consists of the alerts produced by the AMiner \citep{aminer}, Suricata \citep{suricata}, and Wazuh \citep{wazuh} IDSs monitoring the computer networks over the course of several days while various scans and exploits are executed against them.
In each of the 8 scenarios, the same multi-step attack was launched against the respective computer networks with the only significant differences between the scenarios being certain attack parameters such as the duration of the scan attacks.
Based on these differences, we defined a split over the 8 scenarios of AIT-ADS in training, validation and test sets so that the different attributes of the scenarios are evenly distributed across the split.
An overview of the attacks occurring in AIT-ADS in given in Table \ref{tab:attacks}, while a summary of the scenarios of AIT-ADS is provided in Table \ref{tab:scenarios}.

\begin{table}[t]
    \centering
    \footnotesize
    \begin{tabular}{ccrr}
        \hline
        event label & type & duration & number of alerts\\ 
        \hline
        dns\_scan & scan & 0s & 9 \\
        service\_scan & scan & 3s - 20s & 23 - 50 \\
        wpscan & scan & 12s - 52s & 1.5k - 13k \\
        dirb & scan & 11s  - 26min & 4.5k - 428k \\
        webshell\_cmd & exploit & 5s - 2min & 2 - 4 \\
        online\_cracking & exploit & 0s & 2 - 4 \\
        crack\_passwords & exploit & 0s - 33min& 1 - 2\\
        escalated\_sudo\_command & exploit & 5s - 27s & 7 - 41 \\
        attacker\_change\_user & exploit & 0s - 1s & 9 - 17 \\
        dnsteal & exploit & 1d - 4d & 2 - 8\\
        \hline
    \end{tabular}
    \caption{The attacks occurring in AIT-ADS \citep{aitads}.}
    \label{tab:attacks}
\end{table}

\begin{table}[t]
    \centering
    \footnotesize
    \begin{tabular}{rcccccrc}
        \hline
         id & dirb & \makecell{service\\scan} & \makecell{crack\\passwords} & \makecell{dns\\scan} & \makecell{online\\cracking} & duration & split\\
         \hline
         1 & long  & \checkmark & \checkmark & & \checkmark & 4 days & test\\
         2 & long  & \checkmark & \checkmark & & & 4 days & val\\
         3 & short & \checkmark & \checkmark & & & 3 days & test\\
         4 & short & \checkmark & \checkmark & & \checkmark & 3 days & val\\
         5 & short &            & \checkmark & \checkmark & & 5 days & train\\
         6 & short & \checkmark & \checkmark & & & 4 days & train\\
         7 & long  & \checkmark &            & & & 4 days & train\\
         8 & long  & \checkmark & \checkmark & & & 5 days & train\\
         \hline
    \end{tabular}
    \caption{Summary of the 8 scenarios of AIT-ADS \citep{aitads} and the attacks, that do not occur in all of them.}
    \label{tab:scenarios}
\end{table}

There are two issues with this dataset however.
First, it only covers the low alert densities that can be handled well by existing alert grouping solutions already, and, second, it does not contain simultaneous attacks.
Thus, the dataset, in its existing form, is not suitable for our evaluation purposes as it does not allow to evaluate the AlertBERT framework under high alert densities and simultaneously occurring attacks.
For this reason, we developed a data augmentation method which lets us create scenarios with higher noise levels and simultaneously occurring attacks based on the the existing AIT-ADS.
The resulting dataset \textit{AIT Alert Dataset - Augmented} (AIT-ADS-A) allows us to evaluate AlertBERT in settings that are more realistic and significantly degrade the efficacy of existing alert grouping solutions.
Furthermore, AIT-ADS-A has the advantage that one can implement different noise levels in it, which permits us to evaluate AlertBERT under various noise densities and determine their influence on its performance.
The following subsections will outline our data augmentation method, for more details we refer to its documentation\footnote{\url{https://github.com/ait-aecid/AlertBERT/tree/main/aitads_augmented}} in our code repository.

\subsubsection{Augmentation Method}
\label{sec:aug}

To recombine different alert sequences in the most flexible way while retaining all meaningful relations among them, each scenario of AIT-ADS is split into the subsequences corresponding to the respective alert labels.
That is, each full alert sequence is separated in sequences corresponding to the false positive alerts on the one hand, and the alerts which are triggered by each step of the attack chain on the other hand.
The different alert sequences for false positives and attacks can then be freely recombined, through specification of a configuration file, to create new alert sequences which have the same underlying syntactic and semantic structure as the original AIT-ADS.

The elementary unit of an AIT-ADS-A configuration is a \textit{day of alerts}, which can consist of the false positive alerts of several days in AIT-ADS and multiple attacks defined to take place at a certain time during this day.
All the specified alerts of this day will then be assigned timestamps which make them appear to have happened throughout this day and the different sequences are merged according to their timestamps into a single alert sequence representing this day in AIT-ADS-A.
Similarly to AIT-ADS it is also possible in AIT-ADS-A to concatenate multiple such days into a continuous scenario, which is, for example, reasonable if multiple days use noise or attacks from the same scenarios in AIT-ADS.

A simple example of this procedure is illustrated in Figure \ref{fig:augmentation}.
There, on the left side we have two 2-day scenarios called \textquote{A} and \textquote{B} of an original dataset which contains little noise alerts and no simultaneous attacks.
In the middle we find the 7 different alert sequences in which our augmentation method separates the original dataset -- 4 days of noise alerts and 3 attack alert sequences.
On the right side we have two augmented scenarios comprised of these data:
One \textquote{more-noise} scenario consisting of the original scenario \textquote{A} overlaid with the noise from original scenario \textquote{B}.
And one \textquote{simultaneous attacks} scenario consisting of the original scenario \textquote{B} and the scan attack of original scenario \textquote{A}, where the attack sequences are arranged in such a way that both stages of the exploit coincide with the scan attack.

Several design decisions had to be made to ensure that the resulting structure of AIT-ADS-A is both performant and flexible while only allowing for suitable configurations of the data to be made.
Most important among these are:
\begin{enumerate*}[label=(\roman*)]
    \item As AIT-ADS-A uses the true alert groups for data augmentation, it does not dynamically create new configurations during model training to ensure that the training remains completely self-supervised and only requires a finite amount of data.
    \item As it is not possible to subsample a sequence of false positive alerts while ensuring that its structure remains consistent with a real sequence of false positive alerts, it is not possible to adjust noise levels in AIT-ADS-A in a continuous way.
    Instead, the noise level of a configuration is a positive integer and it determines how many noise sequences are overlaid on each day of it.
    \item As the frequency of false positive alerts in AIT-ADS varies considerably over the course of each day of recordings, and merging alerts from different daytimes to coincide would create structures not observed in real data, different sequences of false positive alerts are always merged so that the alerts match in the time-of-day at which they occurred.
\end{enumerate*}

Finally, as our augmentation method allows for the placement of multiple instances of the same attack alerts within a scenario where they should represent different instances of the same attack type, it is also necessary to be able to distinguish these different instances in the evaluation of alert grouping models.
For this reason, our augmentation method also facilitates an augmentation of the alert group labels of AIT-ADS by extending them in a hierarchical manner with 3 levels:
\begin{enumerate*}[label=(\roman*)]
    \item the original alert group labels from AIT-ADS,
    \item the attack stage, and
    \item a unique attack identifier.
\end{enumerate*}
The first level indicates what general type of attack the alert belongs to, i.e.\ in AIT-ADS-A these are just the event labels of AIT-ADS.
The second level is used to distinguish different phases of attacks that can generally occur independently in time -- e.g.\ the two stages of the exploit shown in Figure \ref{fig:augmentation} should be distinct on this level. 
Finally, the third level is a unique identifier which serves to differentiate between different instances of the same attack alerts -- e.g.\ if one inserts the same attack alerts on two different days of a configuration, they are distinguishable on this level.
Using these labels, the goal of the alert grouping problem is to group together all alerts having the same label on all three hierarchy levels, and to measure the alert grouping ability on a high-level label one can calculate the respective macro score over the lower-level labels contained in it.

\subsubsection{Implemented Configurations}

To evaluate the AlertBERT framework, we implemented the following configurations of AIT-ADS-A:
\begin{itemize}
    \item original: This baseline configuration recreates the original AIT-ADS up to the small difference that the false positive alerts of the first day of each scenario are discarded as they contain the training phase of the anomaly detectors of the AMiner IDS, which is not representative for the standard detection mode\citep{aitads}.
    \item simul-attacks: In this configuration, the noise, number of scenarios, and duration is the same as in the \textquote{original} configuration, but the attacks are rearranged so that the \textquote{wpscan} and \textquote{dirb} scans overlap while the remaining attacks occur throughout these scans.
    \item more-noise-1/2/6/11: In this family of configurations the attacks follow the same structure as in the \textquote{original} configuration, but the densities of noise alerts in the data are increased by overlaying false positive alerts from multiple scenarios and days.
\end{itemize}

While the design of the \textquote{original} and \textquote{simul-attacks} configurations is straightforward, the family of configurations \textquote{more-noise-$x$} required a systematic approach to create data augmentations, which allow for a meaningful comparison of alert grouping problems across noise levels.
Fundamentally, in each configuration more-noise-$x$ every day contains $x+1$ days of noise recordings of AIT-ADS, that is the configuration has noise level $x+1$.
In order to keep the noise realistic despite the data augmentation, though, it is necessary to let every day within a scenario of a configurations have a similar distribution of noise, and to let every noise sequence occur equally often in the configuration.
Thus, the increase of noise was implemented by only reducing the number of days per scenario and sharing of noise sequences across scenarios within the train-val-test splits, while the number of scenarios and the attacks belonging to each scenario were left the same.
In this way, all configurations of AIT-ADS-A retain the 8 scenarios of AIT-ADS with their respective attacks, and the train-val-test splits are consistent across all configurations.

Furthermore, while the configuration \textquote{original} contains approx.\ 700.000 noise alerts, and configuration \textquote{more-noise-1} accordingly approx.\ 1.4 million,  the number of noise alerts is capped at around 2.1 million in the configurations with higher noise levels.
This is done to maintain comparability between different configurations as, otherwise, the overall signal/noise ratio in the data would become very low, and the noise would become repetitive and, thus, unrealistic.
To implement this limit, the number of days in the configurations \textquote{more-noise-6} and \textquote{more-noise-11} is reduced by gradually removing the days where no attacks occur until we are left with only one day per scenario in \textquote{more-noise-11}.
As such, the configuration \textquote{more-noise-2/6/11} form an appropriate suite of datasets for alert grouping under high noise levels because all of them contain the same attacks and approximately the same total amount of noise and will, thus, be used in our the evaluation of AlertBERT.
An overview of the configurations is given in Table \ref{tab:noise}.

\begin{table}[t]
    \centering
    \footnotesize
    \begin{tabular}{crrr}
        \hline
        configuration & noise level & number of noise alerts & duration\\
        \hline
        original & 1 & 712.304 & 32 days\\
        simul-attacks & 1 & 712.304 & 32 days \\
        more-noise-1 & 2 & 1.424.608 & 32 days\\
        more-noise-2 & 3 & 2.136.912 & 32 days\\
        more-noise-6 & 7 & 2.205.741 & 14 days\\
        more-noise-11 & 12 & 2.083.900 & 8 days\\
        \hline
    \end{tabular}
    \caption{Implemented configurations of AIT-ADS-A.}
    \label{tab:noise}
\end{table}

\subsection{BERT-Model Implementation}

For our prototype implementation of the embedding-phase of the AlertBERT framework we based our BERT models on the current state-of-the-art ModernBERT \citep{modernbert} architecture.
However, we did not make use of its local-attention mechanism, because we only employed models consisting of few layers.
In addition, we used the timestamps of alerts instead of their relative position in the alert sequence as input for the rotary positional encoding \citep{rope}.
Furthermore, we removed the 25\% of lowest frequencies in the rotary positional encoding as recommended by \citep{barbero2025round, chen2024hope} to increase the stability of the encoding mechanism.
The alert tokenisation, mapping to initial embeddings, and training objective were implemented in the same way as it is discussed in Section \ref{sec:emb} for the case of hard-coded features of alerts.
The threshold to consider tokens as unknown was set to 10 occurrences in the training data based on a visual inspection the token frequency histogram.

The precise hyper-parameters of our models deviating from the ModernBERT \citep{modernbert} implementation are as follows:
Each of the examined models consists of 1 transformer encoder layer with 16 dimensions per attention head and 1, 2, or 4 heads.
The training of the models used context size 4096 and batch size 16, it was conducted with the Adam \citep{kingma2017adammethodstochasticoptimization} optimizer for 60k steps with a linear learning rate schedule and a maximal learning rate of 0.005.
The context size was selected to be as large as possible without negatively impacting the training of the model, and the remaining training parameters were selected based on which yielded the best resulting training objective. 

In the grouping-phase, we use the output embeddings of the BERT-model and apply principal components analysis with dimension 2 as baseline dimensionality reduction before the clustering.
To obtain the alert embeddings for the grouping stage we split the context length of 4096 into a readout window of length 2048 and padding of size 1024 on both ends of the sequence to ensure the stability of the embeddings.

\subsection{Metrics}

To assess the capabilities of alert grouping methods, we compute ROC-curves for the predicted alert groupings with respect to the hyper-parameters of the grouping phase.
This is done by computing the respective true-positive-rate (TPR) and true-negative-rate (TNR) (or, equivalently, the false-positive-rate) for the groupings obtained with various grouping parameters.\footnote{
Positive and negative clustering outcomes are defined based on matching predictions for pairs of data points, cf.\ Section 16.3 of \citep{Manning_Raghavan_Schütze_2008}.
}
This approach allows us to measure the efficacy of an alert grouping method independently of its hyper-parameters at the grouping stage.
Furthermore, it has been shown that ROC-curves, as well as their visual inspection, are an effective evaluation method for density-based clustering that is stable for different data set sizes and cluster imbalances \citep{Jaskowiak2022}.

For the time-delta method, this clustering-based ROC-curve is straightforward to define because the time-delta method has the threshold value $\delta$ as its only grouping parameter, and both the true-positive-rate and true-negative-rate vary monotonically with it.
Thus, it is completely analogous to a standard ROC-curve for a binary classifier.

For the AlertBERT grouping method, it is slightly more complicated to define an appropriate ROC-curve as it has two grouping hyper-parameters, the threshold value $\delta$ and the tuning factor $\theta$ of the time-cosine-metric.
Nonetheless, it is possible to do so in this case as well because the true-positive-rate and true-negative-rate vary monotonically with both parameters.
To generalise the definition of an ROC-curve from varying over one parameter to two parameters, we consider that, in the single parameter case, the definition of an ROC-curve is equivalent to the minimal monotone step function traversing all pairs of TNR/TPR values observed in the results when varying the parameter.
Thus, in the two parameter case, one can simply relax this definition to be the minimal monotone step function enveloping all observed pairs of TNR/TPR values observed in the results when varying both parameters.
In this way, it is possible to take account for the fact that, in the 2 parameter case, not all observed TNR/TPR pairs need to lie on the same step function while retaining the desired property of ROC-curves telling us what is the best achievable true-positive-rate for a given true-negative-rate and vice versa.

To evaluate the results obtained in our experiments, for each examined model, we compute true-positive-rates and true-negative-rates for each hierarchical label occurring in the data and parameters $\delta$ and $\theta$ ranging over the values $a\cdot2^i$ for $i=-7,\dots,13$ and $a\in\{1,1.5\}$.
These parameter ranges were selected as they cover all relevant results observed in our experiments.
Also, we compute the evaluation metrics both including and excluding the noise alerts to differentiate the grouping ability of the examined methods in both, grouping noisy alerts, and grouping only true alerts.
Importantly, we have to note that only the evaluation including the noise alerts is instructive of the results obtained in an unsupervised deployment scenario while the evaluation excluding noise alerts is more indicative of the ability of the grouping methods to separate different attacks, but not directly observable without knowing which alerts are false positives.
To obtain ROC-curves for higher-level labels, for given parameters we aggregate the lower-level TPR and TNR results into macro scores by averaging over all relevant lower-level labels.

\subsection{Results}
\label{sec:results}

\newcommand{\subwidth}{0.4\linewidth}

\newcommand{\simulheads}{2h}
\begin{figure*}[p]
    \centering
    \begin{subfigure}[]{\subwidth}
        \centering
        \includegraphics[width=\linewidth]{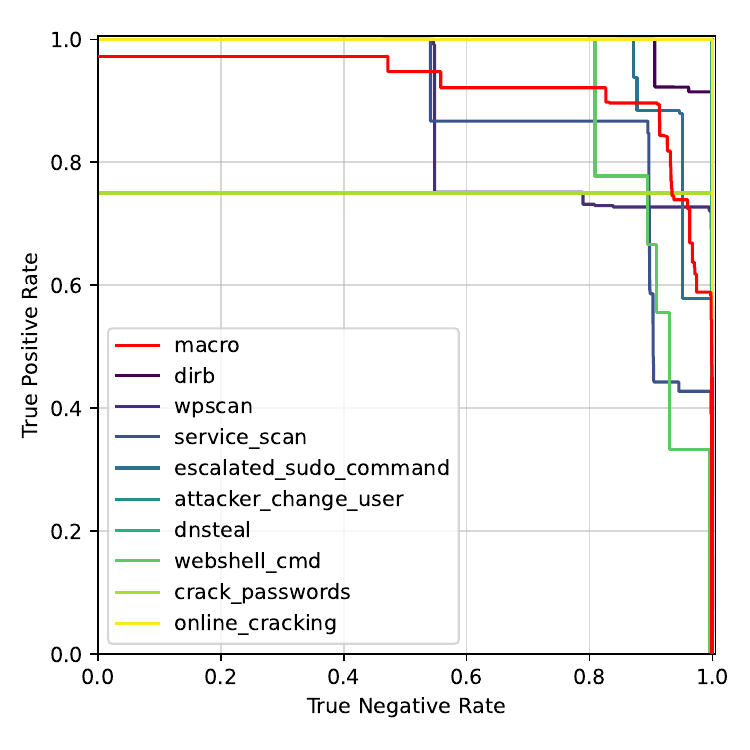}
        \caption{AlertBERT, including noise alerts}
        \label{sub:sim_ab_i}
    \end{subfigure}
    \begin{subfigure}[]{\subwidth}
        \centering
        \includegraphics[width=\linewidth]{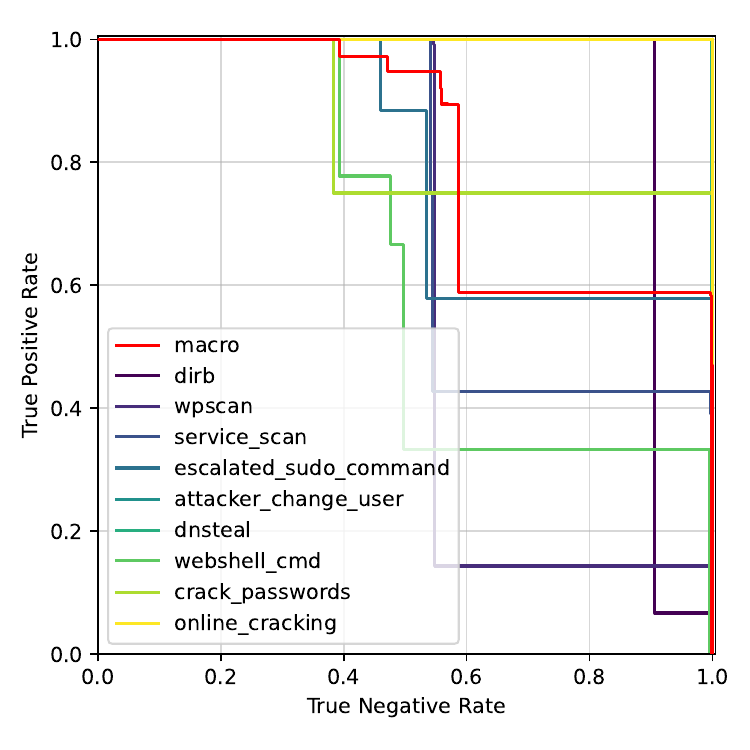}
        \caption{time-delta, including noise alerts}
        \label{sub:sim_td_i}
    \end{subfigure}
    \bigskip
    
    \begin{subfigure}[]{\subwidth}
        \centering
        \includegraphics[width=\linewidth]{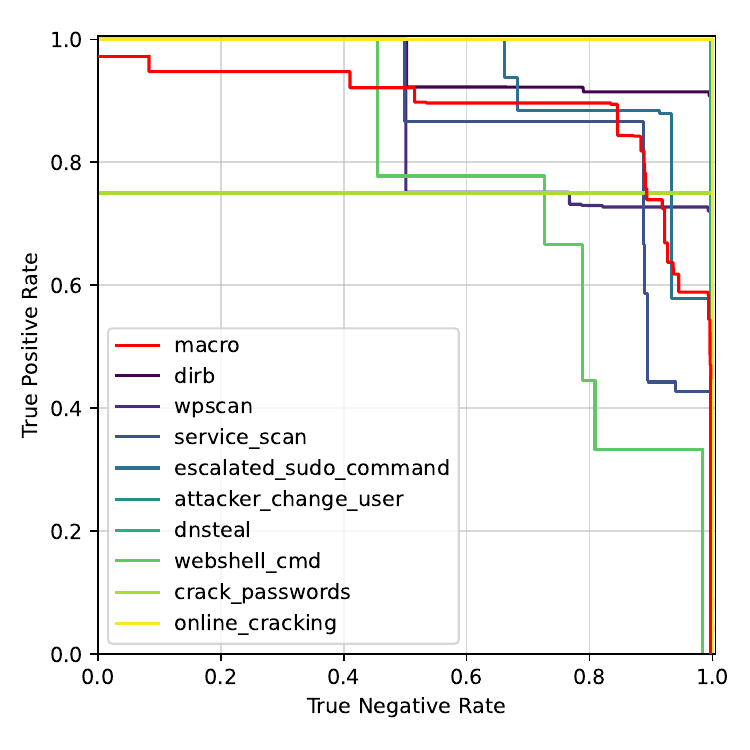}
        \caption{AlertBERT, excluding noise alerts}
        \label{sub:sim_ab_e}
    \end{subfigure}
    \begin{subfigure}[]{\subwidth}
        \centering
        \includegraphics[width=\linewidth]{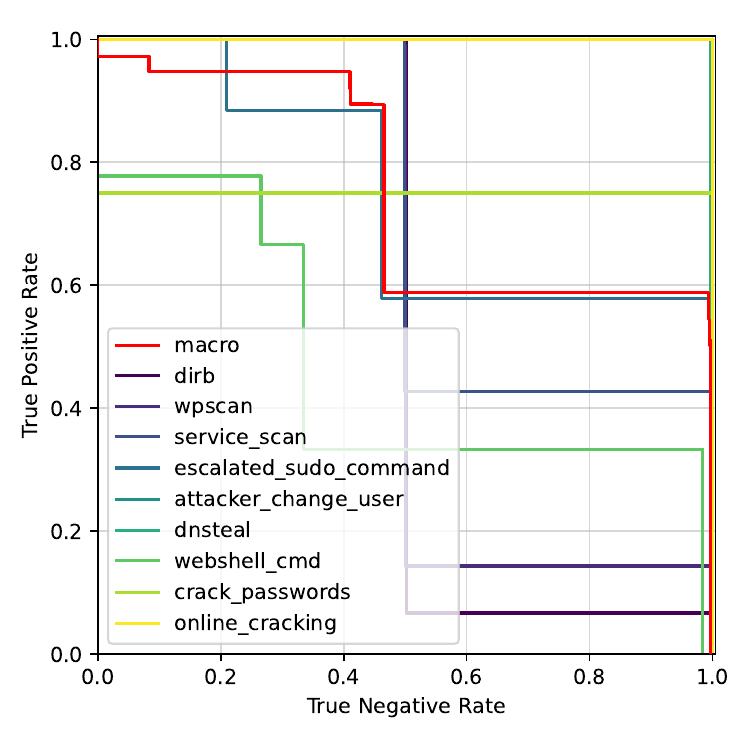}
        \caption{time-delta, excluding noise alerts}
        \label{sub:sim_td_e}
    \end{subfigure}
    \caption{ROC-curves of AlertBERT and time-delta on the \textquote{simul-attacks} configuration of AIT-ADS-A.}
    \label{fig:roc_simul}
\end{figure*}

\begin{table*}[p]
    \renewcommand{\arraystretch}{1.2}
    \footnotesize
    \centering
        \begin{tabular}{cl|r|rrrrrrrrr}
            \hline
            \makecell{noise\\alerts} & method & macro & dirb & wpscan & \makecell{service\\scan} & \makecell{escalated\\sudo\\command} & \makecell{attacker\\change\\user} & dnsteal & \makecell{webshell\\cmd} & \makecell{crack\\passwords} & \makecell{online\\cracking} \\
            \hline
            \multirow{2}*{incl}
            & time-delta &  0.81966 &   0.91226 &   0.61277 &   0.73797 & 0.79401   &   0.99966 & \textbf{0.99906} & 0.63736   & \textbf{0.84582} & \textbf{1.00000} \\
            & AlertBERT & \textbf{0.92585} & \textbf{0.99213} & \textbf{0.88180} & \textbf{0.89503} & \textbf{0.97005} & \textbf{0.99970} & \textbf{0.99906} & \textbf{0.91843} &   0.75000 & \textbf{1.00000} \\
            \hline
            \multirow{2}*{excl} 
            & time-delta &   0.75311 &   0.53551 &   0.57262 &   0.71335 &   0.74158  &   0.99938 & \textbf{0.99718} &   0.46891 & \textbf{0.75000} & \textbf{1.00000} \\
            & AlertBERT &  \textbf{0.88977} & \textbf{0.95829} & \textbf{0.86946} & \textbf{0.88576} & \textbf{0.94037} & \textbf{0.99954} & \textbf{0.99718} & \textbf{0.77461} & \textbf{0.75000} & \textbf{1.00000} \\
            \hline
        \end{tabular}
    \caption{AUC scores of the ROC-curves of AlertBERT and time-delta on the \textquote{simul-attacks} configuration of AIT-ADS-A shown in Figure \ref{fig:roc_simul}.}
    \label{tab:auc_simul}
\end{table*}

\begin{table*}[t]
    \renewcommand{\arraystretch}{1.2}
    \centering
    \footnotesize
        \begin{tabular}{ccl|r|rrrrrrrrr}
            \hline
            \makecell{data\\configuration} & \makecell{noise\\alerts} & method & macro & dirb & wpscan & \makecell{service\\scan} & \makecell{escalated\\sudo\\command} & \makecell{attacker\\change\\user} & dnsteal & \makecell{webshell\\cmd} & \makecell{crack\\passwords} & \makecell{online\\cracking} \\
            \hline
            \hline
            \multirow{4}*{more-noise-2} & \multirow{2}*{incl} 
            &   timedelta & \textbf{0.97274} & 0.99831 & 0.99994 & 0.77760 & 0.99965 & \textbf{1.00000} & \textbf{1.00000} & \textbf{0.89970} & \textbf{0.97454} & \textbf{1.00000}  \\
            & & AlertBERT & 0.93600 & \textbf{0.99994} & \textbf{1.00000} & \textbf{0.82566} & \textbf{0.99999} & \textbf{1.00000} & \textbf{1.00000} & 0.77777 & 0.75000 & \textbf{1.00000}  \\
            \cline{2-13}
            \cline{2-13}
            & \multirow{2}*{excl} 
            &   timedelta & \textbf{0.92625} & \textbf{1.00000} & \textbf{1.00000} & 0.50311 & \textbf{1.00000} & \textbf{1.00000} & \textbf{1.00000} & \textbf{0.77810} & \textbf{0.87507} & \textbf{1.00000}  \\
            & & AlertBERT & 0.91958 & \textbf{1.00000} & \textbf{1.00000} & \textbf{0.62791} & \textbf{1.00000} & \textbf{1.00000} & \textbf{1.00000} & 0.77778 & 0.75000 & \textbf{1.00000}  \\
            \hline
            \hline
            \multirow{4}*{more-noise-6} & \multirow{2}*{incl} 
            &   timedelta & \textbf{0.95749} & 0.99515 & 0.99990 & 0.79425 & 0.99949 & 0.99999 & \textbf{1.00000} & \textbf{0.75481} & \textbf{0.86681} & \textbf{1.00000}  \\
            & & AlertBERT & 0.93645 & \textbf{0.99992} & \textbf{1.00000} & \textbf{0.98581} & \textbf{0.99998} & \textbf{1.00000} & \textbf{1.00000} & 0.69355 & 0.75000 & \textbf{1.00000}  \\
            \cline{2-13}
            \cline{2-13}
            & \multirow{2}*{excl} 
            &   timedelta & 0.91612 & \textbf{1.00000} & \textbf{1.00000} & 0.62736 & \textbf{1.00000} & \textbf{1.00000} & \textbf{1.00000} & 0.61142 & \textbf{0.75034} & \textbf{1.00000}  \\
            & & AlertBERT & \textbf{0.91623} & \textbf{1.00000} & \textbf{1.00000} & \textbf{0.90962} & \textbf{1.00000} & \textbf{1.00000} & \textbf{1.00000} & \textbf{0.66667} & 0.75000 & \textbf{1.00000}  \\
            \hline
        \end{tabular}
    \caption{AUC scores of AlertBERT and time-delta on the \textquote{more-noise-2} and \textquote{more-noise-6} configurations of AIT-ADS-A.}
    \label{tab:auc_less}
\end{table*}

In this section, we present the results of our experiments evaluating AlertBERT on the \textquote{simul-attacks} and \textquote{more-noise-$x$} configurations of AIT-ADS-A.
The embedding models used for these results had 2 attention heads for the case of the \textquote{simul-attacks} configuration and 1 attention head for the \textquote{more-noise-$x$} configurations.
These architectures were selected based on which number of attention heads provided the best results on the respective data configurations.

\subsubsection{Data Configuration \textquote{simul-attacks}}
\label{sec:res_simul}

Figure \ref{fig:roc_simul} displays the ROC-curves obtained for AlertBERT and time-delta on the test set of the \textquote{simul-attacks} configuration of AIT-ADS-A while the respective AUC-scores are shown in Table \ref{tab:auc_simul}.
Each of the plots in Figure \ref{fig:roc_simul} shows the respective ROC-curves obtained for each attack in the data as well as the macro ROC-curve highlighted in red, which summarises the results of all the individual attacks.

Most important in these results, AlertBERT produces AUC-scores superior to those of time-delta in 12 out of 18 examined labels while both methods perform equally well in 5 cases and time-delta beats AlertBERT only on the \textquote{crack passwords} label when including the noise alerts in the evaluation metrics.
While the difference in AUC-scores between AlertBERT and time-delta in the evaluation including the noise alerts generally amounts up to 0.2, with AlertBERT reaching scores greater than 0.88 for all but one label and the highest difference being the \textquote{webshell command} label where AlertBERT scores approx.\ 0.28 better than time-delta, the differences observed for the evaluation of purely the true alerts are even larger.
There, we find that AlertBERT generally outperforms time-delta by a score difference of up to 0.3, with the highest observed difference being 0.423 for the label \textquote{dirb}, while overall AlertBERT has a slightly weaker performance only achieving AUC-scores greater than 0.86 on all but two labels.

Altogether, this situation is reflected in the macro AUC-scores where, including the noise alerts, we find that AlertBERT achieves a very good score higher than 0.925, exceeding time-delta by a margin of more than 0.1.
In the case excluding the noise alerts, on the other hand, we find that AlertBERT still achieves a satisfying macro AUC-score of almost 0.89 while having a larger advantage over time-delta, which lags behind by a score difference of more than 0.13.

To understand these results in more detail, it is also very insightful to examine the respective ROC-curves themselves.
The most prominent features in these plots are clearly the large corners missing in the top right of the ROC-curves obtained from time-delta (to be seen in Figures \ref{sub:sim_td_i} and \ref{sub:sim_td_e}), which are responsible for the worse scores of this method.
This is especially striking as the results located in the top right area of our ROC-plots are the most relevant in practice because these are the only results that achieve an acceptable trade-off between TPR and TNR scores.
The missing corners stem from the fact, that many true positive alerts in AIT-ADS-A have whole-second timestamps and, thus, there are large gaps in the TNR and TPR values of time-delta obtained for $\delta<1$ second and $\delta\geq1$ second.
As can be seen in Figures \ref{sub:sim_ab_i} and \ref{sub:sim_ab_e}, this critical flaw of time-delta is effectively mitigated by the grouping method of AlertBERT as it incorporates the alert embeddings as additional dimensions next to the timestamps.

Furthermore, we can see in the ROC-curves that the score differences among the evaluations including and excluding the noise alerts are mainly due to variations in the true-positive-rates.
There, the evaluation including the noise alerts yields better scores than the evaluation conducted purely on true alerts, while the true-negative-rates are almost the same in both cases.
On the one hand, this behaviour indicates that it is on average easier to attribute pairs of noise alerts to a common group than it is to do so for pairs of true alerts with common origin.
On the other hand, it seems to be equally difficult to separate pairs of alerts with different origins, no matter if these pairs include noise alerts or consist of true alerts stemming from different causes.
As this effect occurs in both time-delta and AlertBERT, although less pronounced in AlertBERT, this observation is an important insight into the nature of the alert grouping problem itself and provides a starting point for further research into improving methods to solve it.

\subsubsection{Data Configuration \textquote{more-noise-$x$}}
\label{sec:res_more}

The AUC-scores measured for AlertBERT and time-delta on the configurations \textquote{more-noise-2} and \textquote{more-noise-6} of AIT-ADS-A can be found in Table \ref{tab:auc_less}.
As one can see there, both AlertBERT and time-delta achieve perfect scores on a large proportion of the examined labels with AlertBERT attaining scores of $1.0$ in 20 of the 36 labels.
However, time-delta scores equally well in 17 of these 20 cases, and among the 19 cases not solved perfectly by time-delta, we can identify two clusters.
On the one hand, there are 9 labels where the differences between the AUC-scores of the two methods are very small $(\leq0.01)$ and where AlertBERT scores better in 7 of the 9 cases.
On the other hand, there are the remaining 10 labels where there is a larger detection gap between the two methods (score differences range from $0.05$ to $0.3$) and where both AlertBERT and time-delta outperform each other in 5 cases respectively.
For the macro AUC-scores, we find that AlertBERT beats time-delta only by a small margin in 1 of the 4 cases, while time-delta scores better than AlertBERT in the other 3 cases and does so with somewhat larger leads of up to $0.04$.

Thus, while the results measured for the non-macro labels slightly favour AlertBERT over time-delta, the reverse holds for the macro results.
Altogether, we view these mixed results not as a testament of the relative capabilities of the examined methods, but rather as a consequence of the limited amount of noise available in the data.
As the results obtained using the \textquote{more-noise-11} configuration of AIT-ADS-A show a more differentiated picture of the abilities of the examined methods, we resort to these data for a more meaningful comparison.

\begin{figure*}[p]
    \centering
    \begin{subfigure}[]{\subwidth}
        \centering
        \includegraphics[width=\linewidth]{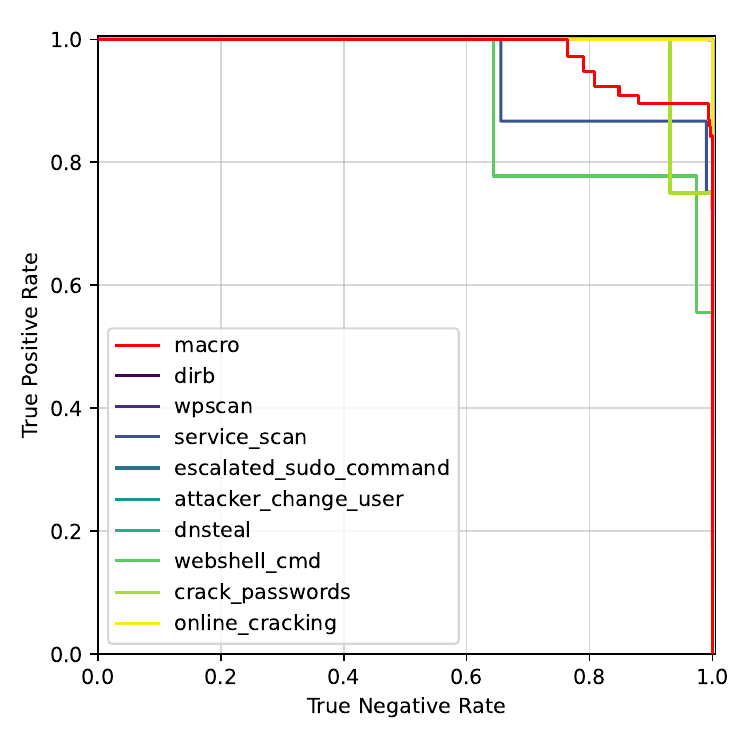}
        \caption{AlertBERT, including noise alerts}
        \label{sub:mor_ab_i}
    \end{subfigure}
    \begin{subfigure}[]{\subwidth}
        \centering
        \includegraphics[width=\linewidth]{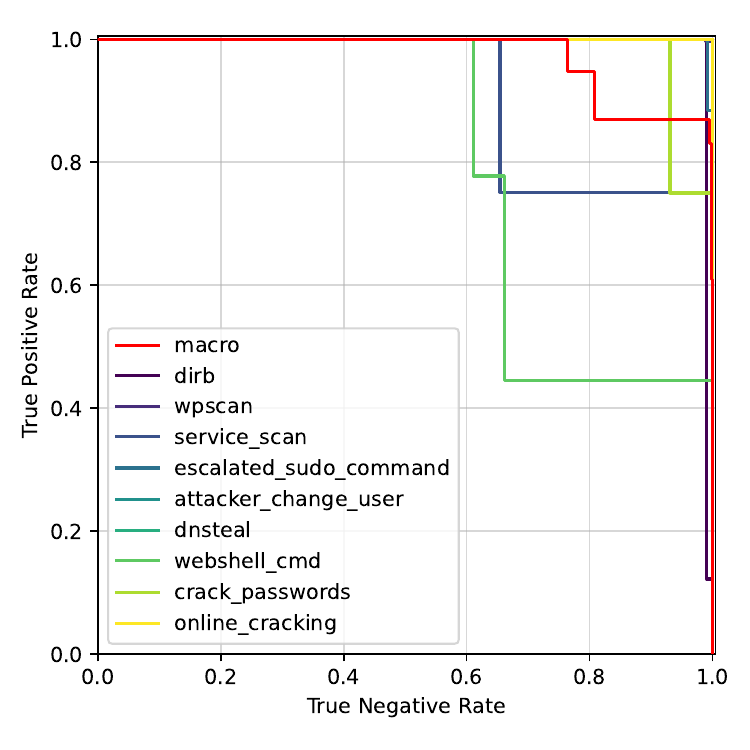}
        \caption{time-delta, including noise alerts}
        \label{sub:mor_td_i}
    \end{subfigure}
    \bigskip
    
    \begin{subfigure}[]{\subwidth}
        \centering
        \includegraphics[width=\linewidth]{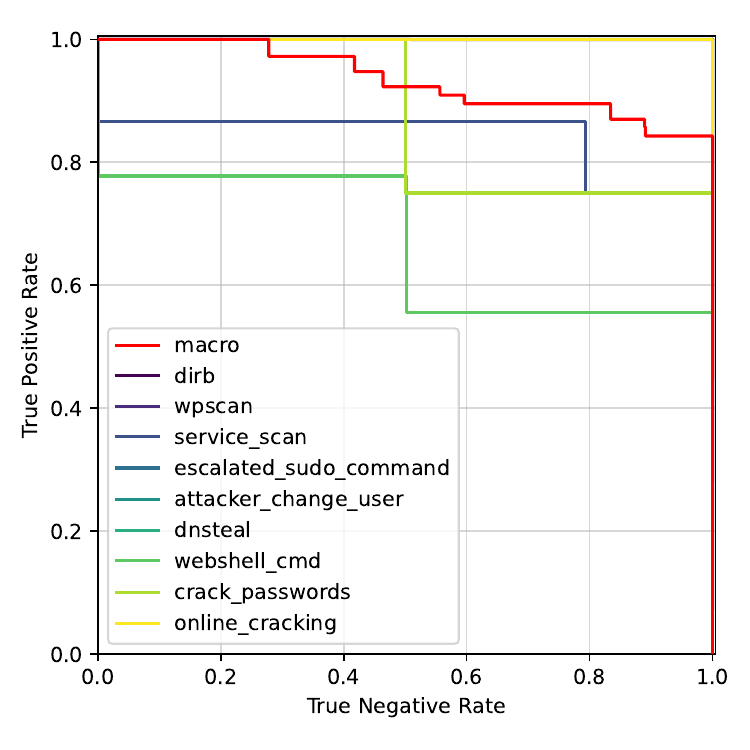}
        \caption{AlertBERT, excluding noise alerts}
        \label{sub:mor_ab_e}
    \end{subfigure}
    \begin{subfigure}[]{\subwidth}
        \centering
        \includegraphics[width=\linewidth]{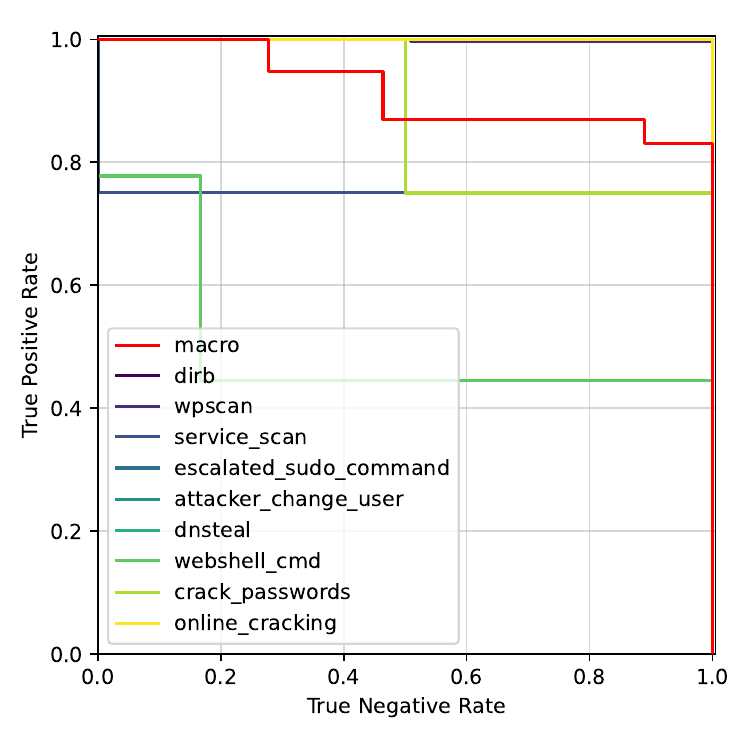}
        \caption{time-delta, excluding noise alerts}
        \label{sub:mor_td_e}
    \end{subfigure}
    \caption{ROC-curves of AlertBERT and time-delta on the \textquote{more-noise-11} configuration of AIT-ADS-A.}
    \label{fig:roc_more}
\end{figure*}

\begin{table*}[p]
    \renewcommand{\arraystretch}{1.2}
    \centering
    \footnotesize
        \begin{tabular}{cl|r|rrrrrrrrr}
            \hline
            \makecell{noise\\alerts} & method & macro & dirb & wpscan & \makecell{service\\scan} & \makecell{escalated\\sudo\\command} & \makecell{attacker\\change\\user} & dnsteal & \makecell{webshell\\cmd} & \makecell{crack\\passwords} & \makecell{online\\cracking} \\
            %\hline
            \hline
            \multirow{2}*{incl} 
            & time-delta &   0.97225 &   0.99159 &   0.99980 &   0.91395 &   0.99897  &   0.99999 & \textbf{1.00000} &   0.80062 & \textbf{0.98271} & \textbf{1.00000} \\
            %\cline{2-12}
            & AlertBERT  & \textbf{0.97942} & \textbf{1.00000} & \textbf{0.99999} & \textbf{0.95292} & \textbf{1.00000} & \textbf{1.00000} & \textbf{1.00000} & \textbf{0.91515} & \textbf{0.98271} & \textbf{1.00000} \\
            %\hline
            \hline
            \multirow{2}*{excl} 
            & time-delta &   0.91618 & \textbf{1.00000} &   0.99856 &   0.75198 &    0.99995 & \textbf{1.00000} & \textbf{1.00000} &   0.50040 & \textbf{0.87507} & \textbf{1.00000} \\
            %\cline{2-12}
            & AlertBERT  & \textbf{0.93347} & \textbf{1.00000} & \textbf{0.99974} & \textbf{0.84333} & \textbf{1.00000} & \textbf{1.00000} & \textbf{1.00000} & \textbf{0.66745} & \textbf{0.87507} & \textbf{1.00000} \\
            \hline
        \end{tabular}
    \caption{AUC scores of the ROC-curves of AlertBERT and time-delta on the \textquote{more-noise-11} configuration of AIT-ADS-A shown in Figure \ref{fig:roc_more}.}
    \label{tab:auc_more}
\end{table*}

In Figure \ref{fig:roc_more} one can see the ROC-curves obtained for AlertBERT and time-delta on the test set of the \textquote{more-noise-11} configuration of AIT-ADS-A while the respective AUC-scores are shown in Table \ref{tab:auc_more}.
First of all, these results indicate that even the amount of noise found in the most extreme configuration of AIT-ADS-A is not enough to properly defeat the time-delta method as it still achieves AUC-scores higher than 0.98 on 13 of the 18 examined labels, and of which it achieves perfect scores on 6 labels.
Nonetheless, AlertBERT is able to improve on this performance as it draws even with time-delta on 8 labels and outperforms it on the remaining 10 labels, achieving perfect scores on a total of 10 labels.

In particular, we highlight the AUC-scores on the 5 labels where time-delta scores less than 0.98 as AlertBERT significantly improves on all these results, except for \textquote{crack passwords} excluding the noise alerts where both methods score equally well.
In the case of the evaluation including the noise alerts these labels are \textquote{service scan} where time-delta attains a score of about 0.914 while AlertBERT reaches about 0.953, and \textquote{webshell cmd} where the score of AlertBERT $(\approx0.915)$ is up by a margin of more than 0.115 from time-delta $(\approx0.8)$.
For the evaluation excluding the noise alerts, we find for \textquote{service scan} that AlertBERT $(\approx0.843)$ surpasses time-delta $(\approx0.752)$ with a score difference of more than 0.09, while for \textquote{webshel cmd} we find the overall largest detection gap of 0.167 between time-delta $(0.5)$ and AlertBERT $(\approx0.667)$.

Based on these results, in the evaluation including the noise alerts both AlertBERT $(\approx0.979)$ and time-delta $(\approx0.972)$ reach excellent macro AUC-scores with a slight advantage for AlertBERT.
In the macro AUC-scores computed excluding the noise alerts, on the other hand, both AlertBERT $(\approx0.933)$ and time-delta $(\approx0.916)$ attain very good scores, but here the advantage of AlertBERT is more apparent.

Finally, in the ROC-plots visible in Figure \ref{fig:roc_more}, we can see that also for the \textquote{more-noise-11} configuration of AIT-ADS-A, the differences in AUC-scores observed between the evaluations including and excluding the noise alerts stem from lower true-negative-rates in the evaluation excluding the noise, while the true-negative-rates are the same for both cases.
This strengthens our conclusion drawn from our observation of this effect in the results on \textquote{simul-attacks} as it appears that this effect is not specific to a data configuration, but seems to have a more general origin.

\section{Discussion}
\label{sec:dis}

This section will relate the results observed in out experiments to the theoretical aspects of the AlertBERT framework and also discuss their limitations and future work.

\subsection{Results}

Across all of our experiments we have seen that the AlertBERT framework provides an improvement over the detection efficacy of the previous state-of-the-art time-delta \citep{Landauer2022Dealing} alert grouping method on challenging datasets.
In particular, our experiments demonstrated that the AlertBERT framework has multiple properties that make it a suitable alert grouping method for deployment in real-world applications where noisy environments and simultaneous attacks are prevalent.

Most important, the incorporation of the outputs of the embedding-phase of the AlertBERT framework as additional dimensions in its alert representation for the grouping-phase (see Section \ref{sec:rep}) is the decisive feature permitting AlertBERT to outperform previous state-of-the-art solutions.
As we have shown with the experiments we discussed in Section \ref{sec:results}, the enhanced alert representation of AlertBERT fulfils two purposes.
On the one hand, the additional embedding dimensions introduced by our framework make the grouping noise-robust.
This claim is supported by our evaluation results obtained on the \textquote{more-noise-11} configuration of AIT-ADS-A discussed in Section \ref{sec:res_more}, which indicate that the additional embedding dimensions allow to avoid collisions among the alert representations in case of high alert density in time.
On the other hand, the embedding dimensions successfully enable AlertBERT to separate simultaneous attacks in the alert grouping.
This claim is supported by the results obtained on the \textquote{simul-attacks} configuration of AIT-ADS-A discussed in Section \ref{sec:res_simul}, which demonstrate that the additional embedding dimensions allow for different alert groups to overlap in time.
Therefore, our experiments show that AlertBERT is capable of producing excellent results on alert grouping tasks entailing high noise levels and simultaneous attacks, which are especially relevant in real-world deployment scenarios.

Moreover, since neither the embedding-phase (described in Section \ref{sec:emb}) of our framework, nor its grouping-phase (described in Section \ref{sec:grp}), require the knowledge of the true causes of alerts, AlertBERT is an self-supervised alert grouping method.
Thus, AlertBERT can be employed directly on top of existing IDSs and is able to produce useful alert groupings without the need for labelled data, which are infeasible to obtain in practice.

Finally, since AlertBERT only requires alerts produced by IDSs (or recordings thereof), and no further contextual information about the systems these IDSs are monitoring as input (cf.\ Section \ref{sec:emb}), it can be operated either in an online or forensic fashion.
This means that AlertBERT can be used for both, analysing security events as they occur, or conducting retrospective analyses, which allows for a flexible use of it.

\subsection{Limitations}

Naturally, there are limitations to our method and to its evaluation.
As main limitation of the AlertBERT framework, we identify the fact that its embedding-phase involves a masked-language-model, which has to be trained on alerts obtained from the computer system that it should work with (or a very similar one).
This means that, to a certain degree, the embedding-phase of AlertBERT does not work \textquote{out of the box} like other alert grouping solutions, but requires effort in the form of configuration and training before initial deployment.

As a further ramification of this, also data drift can pose a problem for the deployment of AlertBERT instances over a prolonged period of time.
This is due to the fact that the structure of the alerts produced by IDSs is inherently unstable over time because it heavily depends on factors that vary over time.
The configuration and version of the used IDSs determine the attributes of alerts and, thus, configuration changes or version updates can add or remove alert attributes, or simply change their names.
Further, also the patterns in which alerts occur are manly influenced by factors fluctuating with time such as the users, devices, and software active in a computer system.
Therefore, it is necessary to take measures to harden AlertBERT against data drift (e.g.\ frequent re-training of the masked-language-model) so that it can maintain its detection capability over extended deployment periods.

As limitations of the evaluation presented here we identify, on the one hand, that we only employed the simplest possible form of alert tokenisation and training objective within our framework.
This has the reason that our evaluation has the purpose of providing a proof-of-concept for the AlertBERT framework, which it does successfully by achieving state-of-the-art results under high-noise and simultaneous-attack conditions.
Thus, there is significant potential to further improve the performance of AlertBERT by using more refined approaches for these steps.
But realising this potential still requires a significant engineering effort to develop these methods and, more importantly, better data, which is the other limiting factor of our evaluation.

The dataset used for our evaluation, AIT-ADS \citep{aitads}, generally has the limitation of only containing recordings of alerts with low levels of noise and no simultaneous attacks.
we tried to mitigate these aspects through data augmentation methods exploiting the entire structure of the dataset as described in Section \ref{sec:aug} and resulting in AIT-ADS-A.
While the \textquote{simul-attacks} configuration of AIT-ADS-A successfully managed to overcome the lack of simultaneous attacks, the \textquote{more-noise-$x$} configurations still suffer from the generally small size of AIT-ADS and do not contain a sufficient amount of noise to accurately reflect many real-world application cases despite having up to the maximal noise level possible under our augmentation method.
For example, \citep{alertfatigue} reports an enterprise cyber security operations centre encountering 137 million alerts over 10 months, which amounts to an average of about 450.000 alerts per day, while the \textquote{more-noise-11} configuration of AIT-ADS-A contains only an average of 260.000 alerts per day.

\subsection{Future Work}

Therefore, as the main topics for future work building on this paper we see, on the one hand, the creation of an alert dataset fixing the limitations of AIT-ADS.
That is an alert dataset that is fully labelled, contains recordings from a large number of computer networks, and which contains simultaneous attacks and realistic levels of noise.
On the other hand, there is a range of possible improvements that can be made to the implementation of AlertBERT evaluated in this paper.
Among these are, for example, the use of more advanced methods for the alert tokenisation step and a training objective appropriate for this, or to utilise the insights gained from the inspection of the ROC-curves presented in Section \ref{sec:results} to further refine the efficacy of the method.

Other than that, possible directions for future work could be to employ methods of \textit{explainable artificial intelligence} (XAI) \citep{MERSHA2025113042} to the masked-language-model part of AlertBERT to learn more about how it perceives the relations among the different alerts, and how this influences the groupings predicted by AlertBERT; or to explore other potential sources for alert embeddings that can be used with the grouping-phase of AlertBERT instead of the the ones produced by the proposed embedding-phase.

\section{Conclusion}
\label{sec:con}

We introduce the AlertBERT framework to tackle the problem of alert grouping in a self-supervised fashion and in application scenarios beyond the capabilities of current methods.
To this end, we propose a two-part theoretical framework to, first, obtain embeddings of alert data by means of a masked-language-model and, then, use these embeddings as part of a density-based clustering method to produce alert groupings.
We evaluate an example implementation of the AlertBERT framework on our AIT-ADS-A dataset involving high levels of noise and simultaneous attacks, and show that it outperforms conventional time-based grouping methods.
Altogether, these features make AlertBERT a versatile and useful tool for cyber security operations and enable analysts to manage computer networks more efficiently.

\section*{Acknowledgments}

Funded by the European Union under the Horizon Europe Research and Innovation programme (GA no.\ 101168144 - MIRANDA) and under the European Defence Fund (GA no. 101121403 - NEWSROOM and GA no.\ 101168092 - ECYSAP EYE). 
Views and opinions expressed are however those of the author(s) only and do not necessarily reflect those of the European Union or the European Commission. 
Neither the European Union nor the granting authority can be held responsible for them.
Co-funded by the Austrian FFG Kiras project ASOC (FO999905301).